\documentclass[preprint]{revtex4}
\usepackage{amsfonts}
\usepackage{amsmath}
\usepackage{amssymb}
\usepackage{graphicx}
\usepackage{booktabs}
\providecommand{\U}[1]{\protect\rule{.1in}{.1in}}

\begin{document}

\title{
{Uncertainty Relations in Thermodynamics of Irreversible Processes on a Mesoscopic Scale}
}

\author{ Giorgio SONNINO}
\affiliation{
Universit{\'e} Libre de Bruxelles (U.L.B.)\\
Department of Theoretical Physics and Mathematics\\
Campus de la Plaine C.P. 231 - Bvd du Triomphe\\
B-1050 Brussels - Belgium\\
Email: giorgio.sonnino@ulb.be
}


\begin{abstract}
Studies of mesoscopic structures have now become a leading and rapidly evolving research field ranging from physics, chemistry, and mineralogy to life sciences. The increasing miniaturization of devices with length scales of a few nanometers is leading to radical changes not only in the realization of new materials but also in shedding light on our understanding of the fundamental laws of nature that govern the dynamics of systems at the mesoscopic scale. On the basis of recent experimental results and previous theoretical research, we investigate thermodynamic processes in small systems in Onsager's region. We show that fundamental quantities such as the total entropy production, the thermodynamic variables conjugate to the thermodynamic forces, and the Glansdorff-Prigogine's dissipative variable may be discretized at the mesoscopic scale. We establish the canonical commutation rules (ccr) valid at the mesoscopic scale. The numerical value of the discretization constant is estimated experimentally. The ultraviolet divergence problem is solved by applying the correspondence principle with Einstein-Prigogine's fluctuations theory in the limit of macroscopic systems.

\end{abstract}

\maketitle


\section{Introduction}
\noindent The physical world can exhibit different behaviors and be described by different laws depending on the scale of observation. This is known as \textit{scale-dependent behavior} or \textit{scale-dependent laws}. The laws that govern macroscopic systems may not necessarily apply directly to mesoscopic or microscopic systems, and vice-versa. Classical physics laws, such as Newtonian mechanics, thermodynamics, and electromagnetism, are typically applicable at the macroscopic scale. These laws have been extensively validated and are effective in describing the behavior of objects and phenomena observable to the naked eye. They provide a good approximation for many everyday situations. However, as we move to smaller scales, such as the mesoscopic and microscopic levels, quantum mechanics becomes increasingly relevant. Quantum mechanics governs the behavior of particles at the atomic, molecular, and subatomic scales. It introduces concepts like wave-particle duality, quantum entanglement, and uncertainty. Quantum effects can also manifest in mesoscopic systems, where collective behavior or quantum coherence comes into play. In certain cases, there can be a smooth transition between classical and quantum behavior as we move from the macroscopic to the microscopic scale. This is known as the \textit{correspondence principle}, where classical laws emerge from quantum mechanics in the limit of large systems and high energies. However, emergent phenomena, as seen at the \textit{mesoscopic scale}, can arise due to the collective behavior of a large number of particles. These phenomena may not have direct counterparts in either classical or quantum mechanics alone, and they often require to establish a new theoretical framework to describe them. A classification, albeit generic, of these three scales can be the following.

\begin{description}
 \item {\bf i)} \textit{Macroscopic Scale}. The macroscopic scale refers to the largest scale of observation, where objects or systems are observed and analyzed on a human scale. It involves the study of phenomena that are visible to the naked eye and can be measured directly. Examples of macroscopic scale observations include the behavior of everyday objects, the motion of planets, or the flow of fluids. At this scale, classical physics laws, such as Newtonian mechanics and thermodynamics, are often applicable;
\item {\bf ii)} \textit{Mesoscopic Scale}. The mesoscopic scale is an intermediate scale between the macroscopic and microscopic scales. It refers to the study of systems or objects that are larger than individual atoms or molecules but smaller than what can be considered macroscopic. The mesoscopic scale often deals with phenomena that arise from the collective behavior of a large number of particles or constituents, exhibiting properties that are not observed at larger or smaller scales. Examples of mesoscopic phenomena include quantum dots, nanoscale devices, and certain biological systems. At this scale, both classical and quantum mechanics may be applicable, but not always, depending on the specific system.
\item {\bf iii)} \textit{Microscopic Scale}. The microscopic scale involves the study of objects or systems at the smallest level of observation, where individual atoms, molecules, or subatomic particles are examined. It deals with phenomena governed by quantum mechanics and involves understanding fundamental particles' behavior, interactions, and properties. Examples of microscopic scale studies include molecular dynamics simulations, atomic structure analysis, or particle physics experiments.
\end{description}

\noindent Currently several research activities are focused on seeking fundamental laws and understanding the behavior of systems at the mesoscopic scale. Scientists and researchers explore mesoscopic phenomena to gain insights into the emergence of novel properties and to bridge the gap between macroscopic and microscopic descriptions. One area of research is \textit{mesoscopic physics}, which investigates systems that are larger than individual atoms or molecules but smaller than macroscopic objects \cite{bachtold}, \cite{cleland1}, \cite{cleland2}, \cite{ekinci}, \cite{lifshitz}. This field aims to understand how collective behavior and quantum effects manifest at the mesoscopic scale. Various experimental techniques, such as nanofabrication and scanning probe microscopy, are employed to study and manipulate mesoscopic systems. Another active research area is \textit{quantum mesoscopics}, which explores mesoscopic systems exhibiting quantum coherence and interference effects \cite{poot}, \cite{seifert}, \cite{jarzynski}, \cite{sagawa}, \cite{berut}, \cite{ribezzi}, \cite{dago}. Researchers investigate phenomena like quantum transport, quantum dots, and mesoscopic superconductivity to uncover the fundamental laws that govern these systems. In this case, advanced experimental techniques and theoretical models rooted in quantum mechanics are employed. Furthermore, research in \textit{condensed matter physics} often delves into mesoscopic systems to investigate emergent phenomena \cite{dinis}, \cite{julicher}, \cite{sheth}, \cite{salle}, \cite{shilo}. This involves studying the behavior of materials with nanoscale dimensions or complex structures, where mesoscopic effects can play a crucial role. For example, the study of topological materials and quantum Hall systems explores the emergence of novel states of matter at the mesoscopic scale. Additionally, studies on nanomechanical vibrational systems have also opened a new direction in the field. Nanomechanical vibrational systems are resonators for mechanical vibrations, and by their nature they are mesoscopic \cite{aspelmeyer}, \cite{schmid}, \cite{steeneken}. Overall, the research activities in these fields aim to uncover fundamental principles and laws that govern the behavior of systems at the mesoscopic scale, bridging the gap between the macroscopic and microscopic worlds and enhancing our understanding of complex phenomena.

\noindent The main objective of this work is to investigate, and consequently try to establish, the thermodynamic laws governing nonequilibrium mesoscopic structures in Onsager's region. We shall see that it is technically possible to propose canonical commutation rules (ccr) for the pair of variables (\textit{time}, \textit{entropic production}) and (\textit{thermodynamic variable}, \textit{conjugate thermodynamic force}). This is motivated by the experimental results: \textit{At the mesoscopic scale, it is not possible to measure the values of transport coefficients and time with infinite precision simultaneously}. Then by extension, \textit{it is also not possible to measure the values of entropy production and time with infinite precision simultaneously}. 

\noindent The manuscript is organized as follows. For easy reference, section~\ref{opf} summarizes the general formalism developed by I. Prigogine and L. Onsager for treating transport processes in thermodynamic systems close to equilibrium. The canonical commutation rules for thermodynamic systems in Onsager's at the mesoscopic scale are proposed in section~\ref{wep}. The numerical value of the \textit{discretization constant} is estimated by using experimental findings in \cite{roldan}. Detailed calculations in the Onsager region, leading to the discretization of the key thermodynamic quantities (such as the total entropy production, the thermodynamic variables conjugate to its thermodynamic forces, and the Glansdorff-Prigogine dissipative quantity) are reported in section~\ref{qtq}. In section~\ref{cep} the \textit{correspondence limit} to Einstein-Prigogine's theory of fluctuations is investigated. In section~\ref{orr} we show that the formalism of the canonical commutation rules does not affect the validity of Onsager's reciprocity relation as long as the flux-force transport relations remain linear. The effect of the volume of the system on the canonical commutation rules is discussed in section~\ref{vccr}. Concluding remarks with a brief introduction on future works being currently finalized can be found in section~\ref{cfw}.

\section{The Onsager-Prigogine Formalism for Systems out of Thermodynamic Equilibrium}\label{opf}
\noindent In order to establish vocabulary, for easy reference, it is useful to recall, very briefly, the formalism used by L. Onsager and I. Prigogine to deal with the thermodynamics of systems out-of-equilibrium. 
\subsection {Prigogine's Theory}
\noindent L. Boltzmann deals with isolated systems, which satisfy the following law: \textit{The total entropy of an isolated system tends to increase over time and reaches thermodynamic equilibrium when the total entropy reaches its maximum value}. I. Prigogine deals with \textit{open systems} and his theory also includes isolated systems as a special case \cite{prigogine1}, \cite{prigogine2}, \cite{prigogine3}, \cite{prigogine4}. Exploiting the additivity property of entropy, Prigogine writes the entropy balance as
\begin{equation}\label{opf1}
d S=d_eS +d_IS
\end{equation} where $dS$ is \textit{the total entropy variation}, $d_eS$ \textit{the entropic flow} i.e., the entropy contribution transmitted reversibly, and $d_I S$ denotes \textit{the entropic production}. Entropy production provides an irreversible contribution and refers to the generation or increase in entropy within a system. By the second law of thermodynamics $d_I S\geq0$. $d S$, $d_eS$ and $d_I S$ have the dimensions of Energy/Temperature. Note that Prigogine writes $d_I S$ and {\it not} $dS_I$ because $d_IS$ \textit {is not an exact differential}. Let us now consider a system characterized by $n$ \textit{degrees of advancement} $\xi_1,\cdots, \xi_n$. The deviations of $\xi_i$ from the values $\xi_\mu^{eq.}$, assumed by the degrees of advance when the system is in local equilibrium, are denoted by ${\bar\alpha}_\mu$, that is ${\bar\alpha}_\mu\equiv\xi_\mu-\xi^{eq.}_i$. Note that the ${\bar\alpha}_\mu$ can also represent the fluctuations of the various thermodynamic quantities (for example, temperature, pressure, etc.).The entropy variation due to the fluctuations is
\begin{equation}\label{opf2}
\Delta S=\Delta_e S +\Delta_IS
\end{equation}
\noindent where
\begin{equation}\label{opf2a}
\Delta S=\int_{\xi^{eq.}}^\xi dS=S(\xi)-S(\xi^{eq.})\ \ ;\ \ \Delta_e S=\int_{\xi^{eq.}}^\xi d_eS\ \ ;\ \ \Delta_I S=\int_{\xi^{eq.}}^\xi d_IS
\end{equation}
\subsubsection{Homogeneous Systems}
\noindent For a non-spatially extended system, the thermodynamic forces are defined as:
\begin{equation}\label{opf3}
{\tilde X}^\mu=\frac{\partial\Delta_I S}{\partial {\bar\alpha}_\mu}\
\end{equation}
\noindent Note that In the Boltzmann case, we get
\begin{equation}\label{opf4}
{\tilde X}^\mu=\frac{\partial\Delta S}{\partial {\bar\alpha}_\mu}=\frac{\partial S}{\partial {\bar\alpha}_\mu}
\end{equation}
\noindent since for an isolated system, we have $\Delta_e S=0$.

\subsubsection{Spatially Extended Systems}
\noindent Let us introduce the thermodynamic forces
\begin{equation}\label{opf5}
x^\mu=\frac{\partial\Delta_I s}{\partial{\bar\alpha}_\mu}
\end{equation}
\noindent where $\Delta_I s$ is the \textit{the entropy production per unit volume} having dimension [$\Delta_I s$]=[Energy]/([Temperature] $\times$[Volume]). Let us now rewrite Eq.~(\ref{opf1}) as
\begin{equation}\label{opf6}
\frac{d S}{dt}=\frac{d_e S}{dt}+\frac{d_I S}{dt}
\end{equation}
\noindent where
\begin{equation} \label{opf7}
\frac{d S}{dt}=\int_V \partial_t s\  dV\quad ; \quad\frac{d_e S}{dt}=-\int_V \nabla\cdot{\bf j}_s\  dV\quad {\rm and} \quad\frac{d_I S}{dt}=\int_V\sigma\ dV
\end{equation}
\noindent Here, ${\bf j}_s$ is the (reversible) \textit{entropy flux} and $\sigma$ the {\it entropy production strength}. $dV$ denotes an infinitesimal spatial volume element and the integrals are performed on the whole volume $V$ occupied by the system. From Eq.~(\ref{opf7}) we get
\begin{equation}\label{opf8}
\partial_t s=-\nabla\cdot{\bf j}_s+{\bar\sigma}
\end{equation}
\noindent Eq.~(\ref{opf7}) also gives
\begin{equation}\label{opf9}
{\bar\sigma}=\frac{\delta}{\delta V}\Bigl(\frac{d\Delta_I S}{dt}\Bigr)=\frac{d}{dt}\Bigl(\frac{\delta\Delta_I S}{\delta V}\Bigr)=\frac{\partial\Delta_Is}{\partial{\bar\alpha}_\mu}\frac{d{\bar\alpha}_\mu}{dt}
=x^\mu j_\mu\quad {\rm with}\quad j_\mu\equiv\frac{d{\bar\alpha}_\mu}{dt}
\end{equation}
\noindent where Eq.~(\ref{opf5}) has been taken into account. $j_\mu$ denotes the \textit{thermodynamic fluxes} conjugate to the thermodynamic force $x^\mu$ \cite{fitts}, \cite{gyarmati}.  In Eq.~(\ref{opf9}) we have adopted the Einstein convention of repeated indices. Unless stated otherwise, this convention will also be adopted in the sequel of this manuscript. Note that at \textit{local equilibrium} (i.e., when $\alpha_\mu=0$), we get
\begin{equation}\label{opf9a}
x^\mu=\frac{\partial\Delta_I s}{\partial \alpha_\mu}\mid_{\alpha_\mu=0}=0
\end{equation}
\noindent To work with entropy production strength, which has dimension [Energy]/([Temperature] $\times$ [time]), we adopt the following definition for the thermodynamic forces, the thermodynamic fluxes, and the thermodynamic variables, respectively:
\begin{equation}\label{opf10}
X^\mu({\bf r},t)\equiv \sqrt{V}x^\mu\quad ; \quad J_\mu({\bf r},t)\equiv \sqrt{V}j_\mu\quad ; \quad \alpha_\mu({\bf r},t)\equiv \sqrt{V}{\bar\alpha}_\mu
\end{equation} 
\noindent with (${\bf r},t$) denoting the space-time. Eq.~(\ref{opf9}) links the entropy production strength with the thermodynamic forces and the conjugate fluxes. To obtain the expression for the entropy production strength solely in terms of the thermodynamic forces, it is necessary to relate the dissipative fluxes to the thermodynamic forces that produce them. These closure relations are called \textit{transport flux-force relations}. The study of these relations is the object of non-equilibrium thermodynamics. The most used transport relations are \cite{balescu1}, \cite{balescu2}, \cite{vidal}, \cite{ottinger}
\begin{equation}\label{opf11}
J_\mu=\tau_{\mu\nu}X^\nu 
\end{equation} 
\noindent with $\tau_{\mu\nu}$ denoting the \textit{transport coefficients}. Note that to perform calculations, the transport coefficients must be written in a dimensionless form (see the forthcoming subsection and \cite{sonnino1}). In terms of the transport coefficients, the local entropy production strength can be brought into the form
\begin{equation}\label{opf12}
\sigma=g_{\mu\nu}X^\mu X^\nu=V g_{\mu\nu}x^\mu x^\nu=V{\bar\sigma}
\end{equation} 
\noindent with $g_{\mu\nu}$ denoting the symmetric piece of the transport coefficients. So, as we wish, $\sigma$ has dimension [Energy]/([Temperature]$\times$[time]) while  ${\bar\sigma}$ has dimension [Energy]/([Temperature]$\times$[time]$\times$[Volume]).

\subsection {Onsager's Theory}\label{of}
\noindent Close to equilibrium, the transport equations of a thermodynamic system are provided by the well-known Onsager theory \cite{onsager1}, \cite{onsager2}. The Onsager transport relations are
\begin{equation}\label{of1}
J_\mu=\tau_{0\mu\nu}X^\nu
\end{equation}
\noindent where $\tau_{0\mu\nu}$ are the transport coefficients. In this equation, the Einstein summation convention on the repeated indexes is adopted. Matrix $\tau_{0\mu\nu}$ can be decomposed into a sum of two matrices, one symmetric and the other skew-symmetric, denoted by $L_{\mu\nu}$ and $f_{0\mu\nu}$, respectively i.e.,
\begin{equation}\label{of2}
\tau_{0\mu\nu}=L_{\mu\nu}+f_{0\mu\nu}\qquad{\rm with}\quad L_{\mu\nu}=L_{\nu\mu}\quad ;\quad f_{0\mu\nu}=-f_{0\nu\mu}
\end{equation}
\noindent The second law of thermodynamics imposes that $L_{\mu\nu}$ be a positive definite matrix. The most important property of Eqs~(\ref{of1}) is that near equilibrium, the coefficients $\tau_{\mu\nu}$ are independent of the thermodynamic forces, i.e.,
\begin{equation}\label{of3}
\frac{\partial\tau_{0\mu\nu}}{\partial X^\lambda}=\frac{\partial L_{\mu\nu}}{\partial X^\lambda}=\frac{\partial f_{0\mu\nu}}{\partial X^\lambda}=0
\end{equation}
\noindent The region where Eqs~(\ref{of1}) and (\ref{of3}) hold, is called {\it Onsager's region} or, the {\it linear region of thermodynamics}. A well-founded microscopic explanation of the validity of the linear phenomenological laws was developed by Onsager in 1931 \cite{onsager1}, \cite{onsager2}. Onsager's theory is based on three assumptions: i) {\it The probability distribution function for the fluctuations of thermodynamic quantities} (Temperature, pressure, degree of advancement of a chemical reaction, etc.) {\it is a Maxwellian} ii) {\it Fluctuations decay according to a linear law} and iii) {\it The principle of the detailed balance} (or the microscopic reversibility) {\it is satisfied}. Onsager showed the equivalence of Eqs (\ref{of1}) and (\ref{of3}) with the assumptions i)-iii) (assumption iii) allows deriving the {\it reciprocity relations} $L_{\mu\nu}=L_{\nu\mu}$). In Onsager's region, the expression for the local entropy production strength reads
\begin{equation}\label{of2}
\sigma=L_{\mu\nu}X^\mu X^\nu=L^{\mu\nu}J_\mu J_\nu
\end{equation}

\subsection {The Minimum Entropy Production Theorem}\label{mep}
\noindent In 1947, I. Prigogine proved the \textit{Minimum Entropy Production Theorem} \cite{prigogine1}, \cite{prigogine2}, concerning the relaxation of thermodynamic systems near equilibrium. This theorem states that:
\vskip 0.2truecm
\noindent {\bf Minimum Entropy Production Theorem}

\textit{A thermodynamic system, near equilibrium, relaxes to a steady state in such a way that the inequality}
\begin{equation}\label{mep1}
\int_V\partial_t\sigma dV\le0
\end{equation}
\noindent \textit{is satisfied throughout the evolution. The inequality is saturated only at the steady state}. 

\noindent It is worth noting that the term $\partial_t\sigma$ consists of two contributions i.e.,
\begin{equation}\label{mep2}
\partial_t\sigma=J_\mu\partial_tX^\mu+X^\mu\partial J_\mu
\end{equation}
\noindent These two contributions are identical (only) in the Onsager region:
\begin{equation}\label{mep3}
J_\mu\partial_tX^\mu=X^\mu\partial J_\mu=2L_{\mu\nu}X^\mu \partial_tX^\nu
\end{equation}
\noindent being $L_{\mu\nu}=L_{\nu\mu}$. Quantity
\begin{equation}\label{mep4}
P=J_\mu\partial_tX^\mu
\end{equation}
\noindent is called the \textit{Glansdorff-Prigogine dissipative quantity} and plays a key role in the thermodynamics of irreversible processes \cite{prigogine5}, \cite{prigogine6}, \cite{degroot}, \cite{sonnino1}, \cite{sonnino2}, \cite{sonnino3}, \cite{sonnino4}, \cite{sonnino5}, \cite{sonnino6}.

\subsection{The Space of the Thermodynamic Forces}\label{ts}
To continue with the formalism it is necessary to define the space where we can perform calculations. For this, we have to specify two quantities: the {\it metric tensor} (denoted by $g_{\mu\nu}$ and the {\it affine connection} (denoted by $\Gamma^\lambda_{\mu\nu}$) \cite{sonnino1}, \cite{sonnino2}, \cite{sonnino3}, \cite{sonnino4}, \cite{sonnino5}, \cite{sonnino6}. 

\noindent {\bf a)} A metric tensor is a central object in the theory; it describes the local geometry of space. The metric tensor is a dimensionless symmetric tensor used to raise and lower the indicative tensors and generate the connections used to determine the field equations, that have to be satisfied by the metric tensor, and to construct the Riemann curvature tensor.

\begin{figure}
\includegraphics[width=4cm]{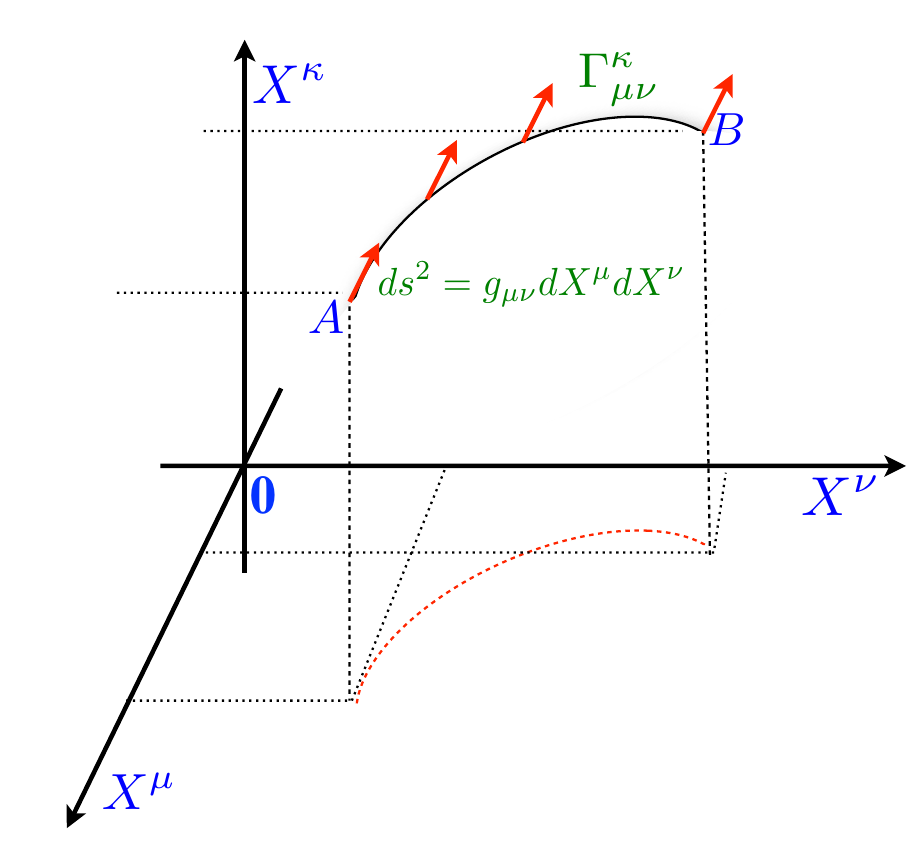}
\caption {\textit{The Thermodynamic Space}. The space is spanned by the thermodynamic forces, the metric tensor is identified with the symmetric piece of the transport coefficients, and the expression of the affine connection is determined by the {\it Universal Criterion of Evolution}.}
\label{fig_Axes}
\end{figure}

\noindent {\bf b)} The curvature of a space can be identified by taking a vector at some point and transporting it parallel along a curve in space-time. An affine connection is a rule that describes how to legitimately move a vector along a curve on the variety without changing its direction. We adopt the following definitions: {\it the space of the thermodynamic forces} (or, simply, {\it the thermodynamic space}) is the space spanned by the thermodynamic forces. The metric tensor and the affine connection are determined by physics. More specifically, the metric tensor is identified with the symmetric piece $g_{\mu\nu}$ of the transport coefficients, and the expression of the affine connection $\Gamma^\kappa_{\mu\nu}$ is determined by imposing the validity of the Glansdorff-Prigogine {\it Universal Criterion of Evolution} \cite{prigogine6}, \cite{sonnino3}, \cite{sonnino6}. Note that, for the second law of thermodynamics, the square (infinitesimal) distance $ds^2=d{\bf s}\cdot d{\bf s}$ is always a nonnegative quantity - see Fig.~\ref{fig_Axes}. 
\noindent In the thermodynamic space, the total entropy production $\sigma_T$, and the minimum entropy production theorem read, respectively
\begin{equation}\label{ts1}
\sigma_T(t)=\frac{1}{\Omega}\int_\Omega g_{\mu\nu}X^\mu X^\nu\sqrt{g}d^nX\qquad ;\quad\frac{1}{\Omega}\int_\Omega\partial_t\sigma\sqrt{L} d^nX\le0
\end{equation}
\noindent with $\Omega$ denoting the (finite) volume of the thermodynamic space where the thermodynamic forces act, and $g$ and $L$ are the determinant of the symmetric piece of the transport coefficients and the determinant of Onsager's matrix, respectively.

\section{The Entropy Production at the Mesoscopic Scale}\label{wep}
\subsection{Preliminary Notions}

\begin{figure*}[htb] 
\includegraphics[width=6cm,height=6cm]{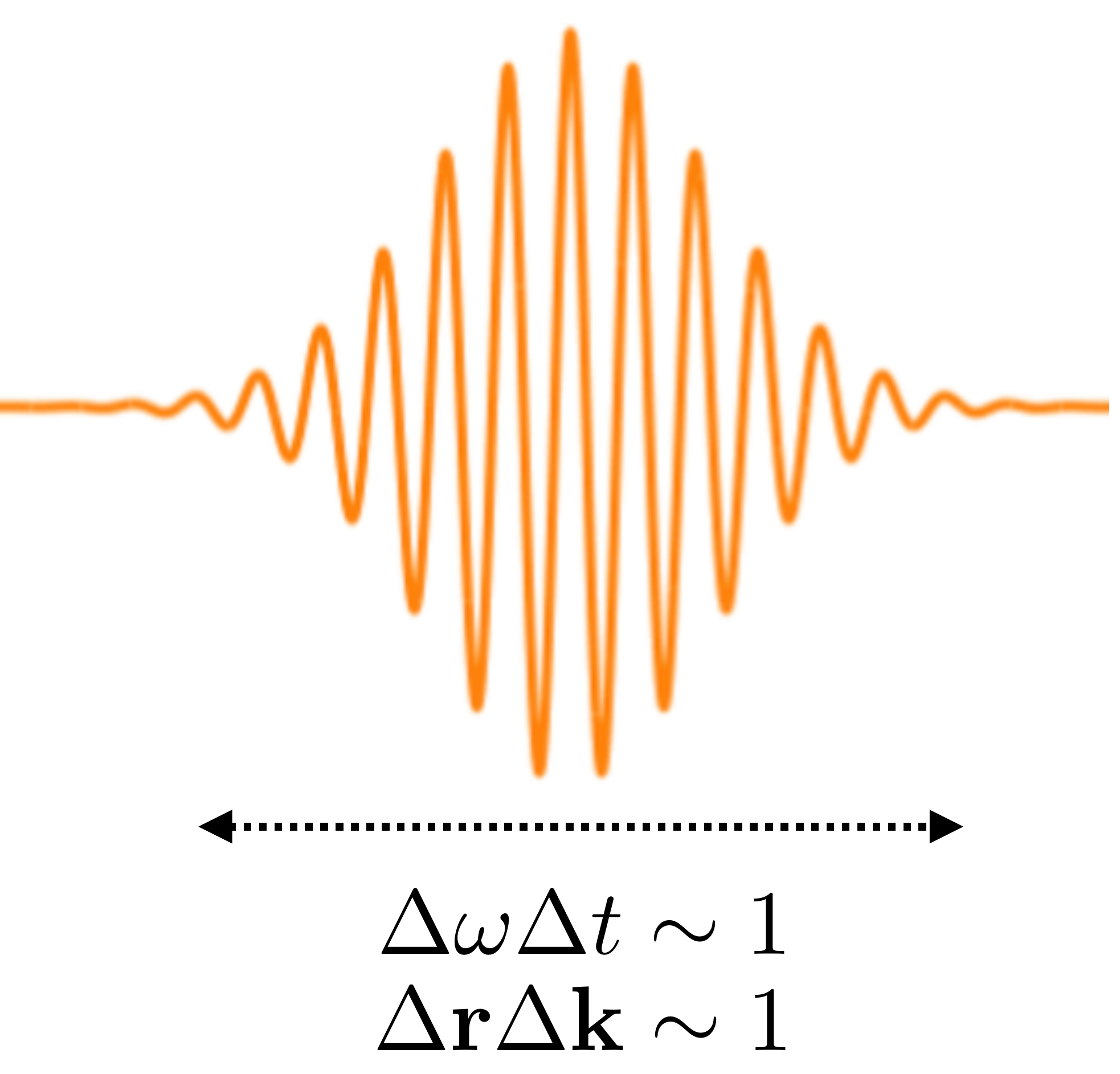}
\caption {\textit{Wavepacket.} A wavepacket is an infinite set of waves with different wave numbers that interfere constructively in a small region and destructively in the rest of space. Fourier's theorem establishes the general property of the wavepackets: $\Delta t\Delta\omega\geq1$ or ${\Delta\bf r}\Delta{\bf k}\geq 1$.}
\label{wp}
\end{figure*}

\noindent In most cases, the study of wave phenomena deals with groups of waves or \textit{wave packets} which are phenomena limited in time and space. For example, the note played by a piano lasts a finite time and the electromagnetic wave produced by the strike of lightning lasts only a fraction of a second. In this case, the function is called a \textit {wave packet} because it is a packet of waves with frequencies/wave numbers clustered around a single value, say $\omega_0$. In these conditions it can be demonstrated that the wave emitted cannot be monochromatic but, on the contrary, it is the sum of several monochromatic waves whose frequency is included in an interval whose measure $\Delta\omega$ is connected to the duration $\Delta t$ of the perturbation. Briefly, a wavepacket is a \textit{localized disturbance} that results from the sum of many different waveforms. If the packed is strongly localized, more frequencies are needed to allow constructive superposition in the region of localization and destructive superposition outside the region.  Fig.~\ref{wp} shows a typical wavepacket. 
\noindent In the $3 + 1$ dimensional space-time (${\bf r},t$), the general form of a real wavepacket $\phi({\bf r},t)$ reads 
\begin{equation}\label{wep1}
\phi({\bf r},t)=\frac{1}{(2\pi)^3}\int_{-\infty}^{+\infty}\left(\phi({\bf k}) e^{i({\bf r}\cdot {\bf k}-\omega({\bf k})t)}+\phi^\star({\bf k}) e^{-i({\bf r}\cdot {\bf k}-\omega({\bf k})t)}\right)d{\bf k}
\end{equation}
\noindent where $\phi^\star({\bf k})$ denotes the complex conjugate of $\phi({\bf k})$. The physical meaning of the above expression is simple: by adding an infinite number of monochromatic waves, chosen appropriately, a localization of the wave perturbation is obtained; or vice-versa, one can imagine that each packet of waves is made up of infinite suitable monochromatic waves. The crucial aspect is that, according to Fourier's theorem, for the amplitudes of any wavepacket, the standard deviations $\Delta t=\sqrt{<t^2>-<t>^2}$ characterizing the width of the spectral distribution in the angular frequency domain, $\Delta\omega$, and the width in time $\Delta t$ are related:
\begin{equation}\label{wep2}
\Delta t\Delta\omega\geq 1
\end{equation}
\noindent This product of the standard deviations equals unity only for the special case of Gaussian-shaped spectral distributions and is greater than unity for all other shaped spectral distributions. This states that the uncertainties with which we can simultaneously measure the time and frequency for the intensity of a given wavepacket are related. Eq. (\ref{wep2}) expresses the celebrated \textit{Fourier's uncertainty principle for frequency-time intensities}.

\subsection{The Entropy Production in Small Systems}
\noindent To avoid misunderstanding, we specify right now that in this approach, entropy production is not considered a wave packet at all. However, as we shall see, Fourier analysis will prove to be a very powerful investigation tool. Entropy production is a macroscopic concept that emerges from the statistical behavior of a large number of particles, and the probabilities and distributions of their microstates. When performing measurements involving a large number of particles and considering the statistical behavior of the system, the resulting macroscopic phenomena, including entropy production, are generally described by continuous variables and distributions. However, it is also true that entropy production arises from collisions among particles in a system. When particles collide and interact, they can lead to changes in the system's microstate configurations, resulting in entropy production. It should be emphasized that all transport coefficients vanish in the absence of collisions. If we perform an experiment at a mesoscopic scale we make observations that reflect the discrete nature of particle interactions. Being the entropic production directly dependent on the transport coefficients (and therefore on the collisions of the particles' system), at the mesoscopic level, the observed entropy production may exhibit a more discrete behavior due to the discrete nature of particle collisions. This is because the discrete nature of individual particle interactions can become more apparent on such a small scale. Fluctuations arise due to the probabilistic nature of individual particle interactions and movements. In smaller systems, these fluctuations have a greater impact on the system's overall behavior.  At a macroscopic scale, the averaging over a large number of particles and the statistical nature of the system smoothes out the microscopic discretization or discrete behavior that may be observed on a smaller scale. Recent experiments show the existence of lower bounds for the rate of entropy production in active fluctuations of the hair-cell bundle by quantifying the irreversibility of stochastic traces obtained from mesoscopic degrees of freedom \cite{roldan}. In particular, it has been observed that, on a mesoscopic scale, the simultaneous measurement of entropy production and time becomes more indeterminate due to the inherent randomness introduced by these fluctuations. Experiments like \cite{roldan} lead to two different (and opposite) interpretations:
\vskip0.2truecm
\noindent {\bf 1)} \textit{The limited number of particles in the system leads to a greater sensitivity to individual events, making it experimentally more challenging to precisely determine entropy production at any specific moment}.

\noindent {\bf 2)} \textit{On a mesoscopic scale, there is a fundamental limit to the accuracy with which the values for time and entropy production can be predicted from initial conditions}.
\vskip0.2truecm
\noindent Here, we shall investigate the second hypothesis. A \textit{quasi-localized} disturbance of entropy production can be obtained using a linear superposition of modes with close mode numbers. So, if the system is subject to "{\textit n}" independent thermodynamic forces, in case of periodic boundary conditions, it is convenient to choose as basis-functions of a \textit{local disturbance} of entropy production strength that of plane waves with the generic wave number ${\bf K}$ compatible with periodicity conditions:
\begin{equation}\label{wep3}
\sigma({\bf X},t)=\frac{1}{(2\pi)^n}\int_{-\infty}^{+\infty}\left(\sigma_{\bf K} e^{i({\bf K}\cdot {\bf X}-\omega_{\bf K}t)}+\sigma^\star_{\bf K} e^{-i({\bf K}\cdot {\bf X}-\omega_{\bf K}t)}\right)d{\bf K}
\end{equation}
\noindent Eq.~(\ref{wep3}) corresponds to a perturbation of the entropy production strength localized in a small region of the thermodynamic space. Notice that the modes are in the space of thermodynamic forces (and not in the ordinary space). Of course, as in ordinary space, even in thermodynamic space it is possible to introduce the \textit {phase velocity} $v_p$ and the \textit {group velocity} $v_g$, defined in the usual way:
\begin{equation}\label{wep5}
v_p=\mid \frac{\omega}{{\bf K}}\mid\qquad;\qquad v_g= \mid\frac{\partial\omega}{\partial{\bf K}}\mid
\end{equation}
\noindent According to the Fourier theorem, we have
\begin{equation}\label{wep5}
\Delta t\Delta\omega\geq 1\quad;\quad \Delta K_\mu\Delta X^\mu\geq 1\quad({\rm no\ summation\ convention\ on}\ \mu;\ \mu=1,\cdots ,n)
\end{equation}
\noindent We may investigate what happens at a \textit{mesoscopic scale} assuming that \textit {in the space of the thermodynamic forces}, the entropy production strength is proportional to the frequency and the thermodynamic variable $\alpha_\mu$, conjugate to the thermodynamic forces $X^\mu$, is proportional to the wave-vector $K_\mu$ i.e., 
\begin{align}\label{wep6}
&\sigma={\slash\!\!\! k}_{\!{B}}\omega\\
&\alpha_{\mu,{\bf K}}={\slash\!\!\! k}_{\!{B}}{\bf K}_\mu\qquad {\rm with} \quad {\slash\!\!\! k}_{\!{B}}\equiv\beta k_B \nonumber
\end{align}
\noindent with ${\slash\!\!\! k}_{\!{B}}$ denoting Boltzmann's constant $k_B$ times a pure number, say $\beta$, undetermined at this stage: ${\slash\!\!\! k}_{\!{B}}\equiv\beta k_B$. Combining Eq.~(\ref{wep5}) with Eq.~(\ref{wep6}) we get
\begin{align}\label{wep7}
&\Delta t\Delta\sigma\geq{\slash\!\!\! k}_{\!{B}}\\
&\Delta\alpha_\mu\Delta X^\mu\geq{\slash\!\!\! k}_{\!{B}}\quad\forall\mu\qquad({\rm no\ summation\ convention\ on}\ \mu;\ \mu=1,\cdots ,n)\nonumber
\end{align}
\noindent These inequalities assert a fundamental limit to how accurately a system's values of certain pairs of physical quantities, such as entropy production strength and time or thermodynamic force and its conjugate thermodynamic variable, can be predicted from initial conditions. In other words, we cannot know with perfect accuracy both the value of the entropy production strength of a system and time; the more we fix the entropy of the system, the less we know about time and vice versa. Similarly, we also have the uncertainty relation between the value of a thermodynamic force and its conjugate thermodynamic variable. The pairs of variables ($t,\sigma$) and ($\alpha_\mu,X^\mu$) may be called complementary variables or {\it canonically conjugate} (in analogy with the quantum mechanics' terminology).

\subsection{The Canonical Commutation Rules}\label{ccr}
At this point, to progress in the formalism, it is necessary to promote the single variables $\sigma$ and $t$, and $X^\mu$ and $\alpha_\mu$ to \textit{operators} (which act on some state space of the system as yet unspecified), separately, imposing a "bind" between them so that their products behave "as we wish". So, at the mesoscopic scale, we write \footnote{This condition finds its justification later. Precisely, thanks to this condition we shall obtain a coherent theory.}:
\begin{align}\label{ccr1}
&[t,\sigma]=i\frac{{\slash\!\!\! k}_{\!{B}}}{2}\\
&[\alpha_{\mu ,{\bf K}},X_{\bf K'}^\nu]=i\frac{{\slash\!\!\! k}_{\!{B}}}{2}\delta_{\mu\nu}\delta_{{\bf K}{\bf K'}}\qquad{\rm with}\quad {\slash\!\!\! k}_{\!{B}}=\beta k_B
\nonumber
\end{align}
\noindent with $[\cdots]$ denoting the \textit {commutator} between two operators: $[A,B]=AB-BA$ and $\delta_{\mu\nu}$ the Kronecker delta, respectively \footnote{Note that the dimensions of the expressions~(\ref{ccr1}) are correct. Indeed, the dimension of Boltzmann's constant is [$k_B$]=[Energy]/[Temperature]. The dimension of the quantity $[t,\sigma]$ is  [time]$\times$ [Energy]/([Temperature]×[time])= [Energy]/[Temperature]=[$k_B$]. The dimension of the quantity $[\alpha_\mu,X^\nu$] is [time]$\times$[$\sigma^{1/2}$]$\times$ [$\sigma^{1/2}$]=[$k_B$].}. In \cite{roldan} we find several examples of experimental traces for the tip position of different mechano-sensory hair bundles as a function of time. The authors estimated the local irreversibility measure obtained from single $30\ sec.$ recordings of the oscillations shown in these examples. The sampling rate was $\omega= 2.5 kHz$. From these experiments, we can estimate (albeit approximately) the numerical value of $\beta$. We find $\beta\sim 1.2\times 10^{-8}$ so, ${\slash\!\!\! k}_{\!{B}}\sim 1.6\times 10^{-31}\ J/K$.

\section{Discretization of the Key Thermodynamic Quantities of a Small System in Onsager's Region}\label{qtq}
In this first work, we shall limit ourselves to discretizing the entropic production of a small system in the Onsager region (i.e., in a region outside the thermodynamic equilibrium but still sufficiently close to it). In this situation, as mentioned in section~\ref{opf}, the metric of the thermodynamic space is flat (i.e., it does not depend on the thermodynamic forces) and the metric's coefficients are the elements of Onsager's matrix $L_{\mu\nu}$. 

\subsection{Discretization of the Total Entropy Production Strength}\label{qep}
The \textit{total entropy production strength} is
\begin{equation}\label{qep1}
\sigma_T(t)=\frac{1}{\Omega}\int_\Omega\sigma({\bf X},t)\sqrt{L}d^nX=\frac{1}{\Omega}\int_\Omega L_{\mu\nu}X^\mu X^\nu\sqrt{L}d^nX
\end{equation}
\noindent The local entropy production strength can be split into two contributions
\begin{equation}\label{qep2}
\sigma({\bf X},t)=L_{\mu\nu}X^\mu X^\nu=\frac{1}{2}L_{\mu\nu}X^\mu X^\nu+\frac{1}{2}L^{\mu\nu}J_\mu J_\nu
\end{equation}
\noindent Let us now perform the following linear coordinate transformation \footnote{Note that linear transformations of coordinates are allowed because this class of transformations belongs to the TCT-group \cite{sonnino4}.}:
\begin{equation}\label{qep3}
{X'}^\lambda=A^\lambda_\kappa X^\kappa,\quad{\rm with}\ \ A^\mu_\nu\ {\rm such\ that}\ \ A^\alpha_\lambda L^{\lambda\kappa}A^\beta_\kappa={\rm I}^{\alpha\beta}
\end{equation}
\noindent with ${\rm I}^{\alpha\beta}$ denoting the Identity matrix. So, after transformation, $\sqrt{L}\rightarrow 1$. Notice that, since the matrix $L_{\mu\nu}$ is a positive definite matrix, there exists always a matrix $A^\mu_\nu$, which satisfies condition~(\ref{qep3}). After transformation we get 
\begin{align}\label{qep4}
&\sigma_T(t)=\frac{1}{{\Omega}'}\int_{\Omega'} \sigma({\bf X}',t)d^nX'\qquad{\rm with}\\
&\sigma({\bf X}',t)=\frac{1}{2}\sum_{\mu=1}^n\left(\mid X^{'\mu}\mid^2+\mid J'_\mu\mid^2\right)\nonumber
\end{align}
\noindent In the space of the thermodynamic forces, the Fourier expansion in a finite box of volume $\Omega$ of the thermodynamic fluxes reads
\begin{equation}\label{qep5}
J'_\mu({\bf X}',t)=\sum_{\bf K}\left(J_{\mu,{\bf K}}(t) e^{i({\bf K}\cdot {\bf X'})}+J^\star_{\mu,{\bf K}}(t) e^{-i({\bf K}\cdot {\bf X'})}\right)
\end{equation}
\noindent where
\begin{equation}\label{qep6}
J_{\mu,{\bf K}}(t) =i\omega_{\bf K}\alpha_{\mu,{\bf K}}(t)\qquad{\rm and}\qquad J^\star_{\mu,{\bf K}}(t) =-i\omega_{\bf K}\alpha^\star_{\mu,{\bf K}}(t)
\end{equation}
\noindent Let us now evaluate the total entropy production strength $\sigma_T(t)$. This task is easily accomplished by taking into account the orthogonality of the plane waves, assumed normalized in a box of volume $\Omega$. We begin by considering the second contribution:
\begin{align}\label{qep7}
&\frac{1}{2\Omega'}\int_{\Omega'}\sum_{\mu=1}^n\mid J_\mu'\mid^2d^nX'=\\
&\frac{1}{2\Omega'}\sum_{\mu=1}^n\sum_{{\bf K},{\bf K}'}\int_{\Omega'}\!\!\left(J_{\mu,{\bf K}}(t) e^{i({\bf K}\cdot {\bf X}')}\!+\!J^\star_{\mu,{\bf K}}(t) e^{-i({\bf K}\cdot {\bf X}')}\right)\!\!
\left(J_{\mu,{\bf K}'}(t) e^{i({\bf K'}\cdot {\bf X}')}\!+\!J^\star_{\mu,{\bf K}'}(t) e^{-i({\bf K'}\cdot {\bf X}')}\right)d^nX'\nonumber
\end{align}
\noindent The orthogonality relation reads
\begin{equation}\label{qep7a}
\frac{1}{\Omega}\int_\Omega e^{i{\bf K}\cdot{\bf X}} e^{-i{\bf K'}\cdot{\bf X}}d^nX=\delta_{{\bf K}{\bf K}'}
\end{equation}
\noindent 
\noindent Additionally, by looking at Eq.~(\ref{qep6}), we have to take into account that $J_{\mu,{\bf K}'}=J^\star_{\mu,{-\bf K}'}$. So, upon integration over the volume, we get two terms with ${\bf K}={\bf K}'$ and two terms with ${\bf K}=-{\bf K}'$, finally giving four equivalent terms in all. Similar calculations apply to the first contribution $\int_{\Omega'}\mid X^{'\mu}\mid^2 d^nX'$ (in this case, $X^\mu_{\bf K}={X^{\mu}}^\star_{\!\!\!\!-{\bf K}'}$). The final result is
\begin{equation}\label{qep8}
\sigma_T(t)=2\sum_{\mu=1}^n\sum_{\bf K}\left(\mid X^\mu_{\bf K}(t)\mid^2+\mid J_{\mu,{\bf K}}(t)\mid^2\right)=2\sum_{\mu=1}^n\sum_{\bf K}\left(\mid X^\mu_{\bf K}(t)\mid^2+\omega_{\bf K}^2\mid \alpha_{\mu,{\bf K}}(t)\mid^2\right)
\end{equation}
\noindent where Eqs~(\ref{qep6}) has been taken into account. This expression is identical to the Hamiltonian of a harmonic oscillator if we set the value of the $mass=4$ and we identify the following terms: \textit{position}$\rightarrow\alpha_{\mu,K}$, \textit{momemtum}$\rightarrow 2X^\mu_K$, and \textit{frequency}$\rightarrow\omega_K$ (see, for example, \cite{griffiths}, \cite{liboff}). So, we first define two new dimensionless operators ${\tilde X}^\mu_K$ and ${\tilde\alpha}_{\mu,K}$, as follows:
\begin{equation}\label{qep9}
{\tilde X}^\mu_{\bf K}=\sqrt{\frac{2}{{\slash\!\!\! k}_{\!{B}}\omega_{\bf K}}}X^{\mu}_{\bf K}\qquad; \qquad
{\tilde\alpha}_{\mu,{\bf K}}=\sqrt{\frac{2\omega_{\bf K}}{{\slash\!\!\! k}_{\!{B}}}}\alpha_{\mu,{\bf K}}
\end{equation}
\noindent In terms of these new variables, the expression for $\sigma_T(t)$ reads
\begin{equation}\label{qep10}
\sigma_T(t)=\sum_{\mu=1}^n\sum_{\bf K}{\slash\!\!\! k}_{\!{B}}\omega_{\bf K} \left(\mid {\tilde X}^\mu_{\bf K}\mid^2+\mid {\tilde \alpha}_{\mu,{\bf K}}\mid^2\right)
\end{equation}
\noindent As for the case of the harmonic oscillator, we assume the validity of the following \textit {Canonical Commutation Rules} (ccr):
\begin{equation}\label{qep11}
[{\tilde \alpha}_{\mu,{\bf K}}, {\tilde X}^\nu_{\bf K'}]=i\delta_{\mu\nu}\delta_{{\bf K}{\bf K'}}\qquad {\rm so}\quad [\alpha_{\mu,{\bf K}},X^\nu_{\bf K'}]=i\frac{{\slash\!\!\! k}_{\!{B}}}{2}\delta_{\mu\nu}\delta_{{\bf K}{\bf K'}}
\end{equation}
\noindent The two operators of \textit {creation} "${\rm a}_{\bf K}^{(\mu)+}$" and \textit {destruction} "${\rm a}_{\bf K}^{(\mu)}$" can be introduced and defined as usual (see, for example, \cite{weinberg}, \cite{maiani}):
\begin{align}\label{qep12}
&{\rm a}_{\bf K}^{(\mu)}=\frac{1}{\sqrt{2}}\left({\tilde \alpha}_{\mu,{\bf K}}+i{\tilde X}^\mu_{\bf K}\right) \\
&{\rm a}_{\bf K}^{(\mu)+}=\frac{1}{\sqrt{2}} \left({\tilde \alpha}_{\mu,{\bf K}}-i{\tilde X}^\mu_{\bf K}\right) \nonumber
\end{align}
\noindent so $[{\rm a}_{\bf K}^{(\mu)}, {\rm a}_{\bf K'}^{(\mu')+}]=\delta_{\mu\mu'}\delta_{{\bf K}{\bf K'}}$. We finally get the discretization of the total entropy production strength in Onsager's region
\begin{equation}\label{qep13}
{\pmb\sigma}_T(t)=\sum_{\mu=1}^n\sum_{\bf K} {\slash\!\!\! k}_{\!{B}}\omega_{\bf K}\left({\rm a}_{\bf K}^{(\mu)+}{\rm a}_{\bf K}^{(\mu)}+\frac{1}{2}\right)
\end{equation}
\noindent Previous expression may also be written in terms of the \textit{number operator} ${\bf n}_{\bf K}^{(\mu)}\equiv {\rm a}_{\bf K}^{(\mu)+}{\rm a}_{\bf K}^{(\mu)}$:
\begin{equation}\label{qep14}
{\pmb\sigma}_T(t)=\sum_{\mu=1}^n\sum_{\bf K} {\slash\!\!\! k}_{\!{B}}\omega_{\bf K}\left({\bf n}_{\bf K}^{(\mu)}+\frac{1}{2}\right)
\end{equation}
\noindent To sum up, \textit {in the space of the thermodynamic forces the total entropy production strength behaves as the sum of "${\bf K}$" (discretized) independent one-dimensional harmonic oscillators, each oscillating with frequency $\omega_{\bf K}$}. 

\vskip0.2truecm
\noindent \textit{\bf\textit{Comments}}

\noindent {\bf i)} As we shall see in section~\ref{cep}, the constant term
\begin{equation}\label{qep15}
\frac{1}{2}\sum_{\mu=1}^n\sum_{\bf K} {\slash\!\!\! k}_{\!{B}}\omega_{\bf K}
\end{equation}
\noindent corresponds to the entropy production produced by fluctuations very close to the \textit{thermodynamic equilibrium} or to the \textit{local non-equilibrium state}. In principle, this additional term is divergent. This problem will be solved in the forthcoming section by applying the \textit{correspondence principle} which states that in the limit of large (macroscopic) systems, our predictions must coincide with those of Einstein-Prigogine's fluctuations theory.

\noindent {\bf ii)} The discretized total entropy production strength does not depend explicitly on time, so the entropy production strength of mode {\bf K} is \textit{stationary}, as are the eigenfunctions of the operator, which are also stationary in time. This \textit{unit of information of entropic production strength} may briefly be referred to as \textit{TUI} (\textit{Thermodynamic Unit of Information}). Also, being
\begin{equation}\label{qep16}
[{\pmb\sigma}_T,{\bf n}_{\bf K}^{(\mu)}]=0
\end{equation}
\noindent the number of \textit{TUI} in each mode for the total entropy production strength remains constant over time.

\noindent {\bf iii)} Given the form of total entropy production strength, all the results found for the oscillator remain valid. In particular, the states of the total entropic production strength are of the type $\mid n_1,n_2,\cdots>$ with $n_1,n_2,\cdots$ positive integers. These states are obtained from the \textit{vacuum state} (corresponding to the thermodynamic equilibrium or to the local non-equilibrium steady state) through the application of the creation operator, and there is no limit to the population of the modes.

\subsection{Probabilistic Treatment of the Entropy Production}\label{pep}
\noindent Entropy production is a statistical quantity that characterizes the behavior of a system over time. It involves considering the fluctuations and probabilities associated with different trajectories and events. In many nonequilibrium systems, the dynamics and behavior of the system involve probabilistic transitions between different states or configurations. So, the probabilistic mathematical framework is well-suited for studying entropy production because it captures the stochastic nature of nonequilibrium systems and provides a means to analyze and quantify the statistical properties and fluctuations associated with entropy production. The operators' formalism introduced in the previous section lends in a natural way to a probabilistic treatment of entropy production. Eq.~(\ref{qep14}) shows that the total entropy production is a self-adjoint operator having eigenvectors of the type:
\begin{equation}\label{pep1}
\mid n_1, n_2,\cdots,n_\kappa,\cdots>
\end{equation}
\noindent for which we have
\begin{equation}\label{pep2}
{\bf n}_{\bf K}^{(\mu)}\mid n_1, n_2,\cdots,n_\kappa,\cdots>=n_\kappa^{(\mu)} \mid n_1, n_2,\cdots,n_\kappa,\cdots>
\end{equation}
\noindent We interpret the eigenvector Eq.~(\ref{pep1}) as the state with $n_1$ \textit{TUI} with mode $K_1$, $n_2$ \textit{TUI} with mode $K_2$, $\cdots$, $n_\kappa$ \textit{TUI} with mode $K_\kappa$, $\cdots$. The eigenvalue $\sigma_T$ of the operator ${\pmb\sigma}_T$ reads 
\begin{equation}\label{pep3}
\sigma_T=\sum_{\bf K}\sum_{\mu=1}^n\left(n^{(\mu)}_\kappa+\frac{1}{2}\right){\slash\!\!\! k}_{\!{B}}\omega_{\bf K}
\end{equation} 
\noindent In accordance with our expectation, Eq.~(\ref{pep3}) shows that the eigenvalues of the entropy production $\sigma_T$ can never be negative. Since a self-adjoint operator has a complete orthonormal system of eigenvectors, we may identify the space of states of the total entropy production strength as the one generated by all vectors of the type Eq.~(\ref{pep1}). In this formalism, Eq.~(\ref{pep3}) may be written as
\begin{equation}\label{pep4}
\sigma_T=<n_1, n_2,\cdots,n_\kappa,\cdots\mid{\pmb\sigma}_T\mid n_1, n_2,\cdots,n_\kappa,\cdots>=\sum_{\bf K}\sum_{\mu=1}^n\left(n^{(\mu)}_\kappa+\frac{1}{2}\right){\slash\!\!\! k}_{\!{B}}\omega_{\bf K}
\end{equation} 
\noindent So, the probabilistic interpretation related to entropic production strength is the following: 

\noindent \textit{The probabilities $p_\kappa$ that $n_\kappa$ \textit{TUI} of value ${\slash\!\!\! k}_{\!{B}}\omega_K$ are detected in a fixed eigenstate, is described by the coefficients of the development of the state on the complete orthonormal system of the eigenvectors of ${\pmb\sigma}_T$}.

\noindent The above somewhat brings the entropy production formalism presented here closer to the one currently in use, which is also based on a probabilistic approach.

\subsection{Discretization of the Thermodynamic Variable Conjugate to the Thermodynamic Force}\label{qf}
\noindent We have already noted that, in Thermodynamic Field Theory, a thermodynamic flux $J_\mu$ that is conjugate to the thermodynamic force $X^\mu$ can be seen as the \textit{covariant aspect of the same quantity} as the two variables are related each with other by the metric tensor $g_{\mu\nu}$ according to the relation $J_\nu=g_{\mu\nu}X^\mu$. In other words, $X^\mu$ and $J_\mu$ are two different aspects of a unique entity. For this reason, these two quantities \textit {commute}. A different thing is the relationship between the thermodynamic force $X^\mu$ and the thermodynamic variable $\alpha_\mu$ conjugate to it; these two quantities \textit {do not commute}. We can easily check that also the quantity $\alpha_\mu$ is discretized according to the relation (\ref{wep6}) reported in section~\ref{wep}. In the thermodynamic space, the local entropy production strength reads. 
\begin{equation}\label{qf1}
\sigma({\bf X},t)=\sum_{\bf K}\left(\sigma_{\bf K}(t) e^{i({\bf K}\cdot {\bf X})}+\sigma^\star_{\bf K}(t) e^{-i({\bf K}\cdot {\bf X})}\right)
\end{equation}
\noindent Then,
\begin{equation}\label{qf2}
J_\mu=\frac{\partial\sigma}{\partial X^\mu}=i\sum_{\bf K}{\bf K}_\mu\left(\sigma_{\bf K}(t) e^{i({\bf K}\cdot {\bf X})}-\sigma^\star_{\bf K}(t) e^{-i({\bf K}\cdot {\bf X})}\right)
\end{equation}
\noindent On the other hand, we also have 
\begin{align}\label{qf3}
&J_\mu=\frac{d\alpha_\mu}{dt}\qquad {\rm so}\\
&\alpha_\mu({\bf X},t)=i\sum_{\bf K}\omega_{\bf K}\left(\alpha_{\mu,{\bf K}}(t) e^{i({\bf K}\cdot {\bf X})}-\alpha^\star_{\mu,{\bf K}}(t) e^{-i({\bf K}\cdot {\bf X})}\right)\nonumber
\end{align}
\noindent By comparing Eq.~(\ref{qf2}) with Eq.~(\ref{qf3}) (or, equivalently, by using the orthogonality relations), we get
\begin{equation}\label{qf4}
\alpha_{\mu,{\bf K}}\omega_{\bf K}=\sigma_{\bf K}{\bf K}_\mu
\end{equation}
\noindent or
\begin{equation}\label{qf5}
\alpha_{\mu,{\bf K}}={\slash\!\!\! k}_{\!{B}}{\bf K}_\mu
\end{equation}
\noindent where Eq.~(\ref{pep4}) has been taken into account. As an example, let us examine a system where "$n$" chemical reactions take place simultaneously. We want to investigate the case where small fluctuations from thermodynamic equilibrium occur \footnote{For simplicity we shall examine the argument by using the degree of advancements as parameters. The generalization is trivial.}. The total entropy production due to the fluctuations $\alpha_\mu$ reads
\begin{equation}\label{qf6}
\Delta_IS=\int_{\pmb{\xi}_{eq.}}^{\pmb\xi} d_IS=\sum_{\mu=1}^n\int_{{{\pmb\xi}_{eq.}}}^{\pmb\xi}\frac{A^\mu}{T}d{\pmb\xi}\qquad{\rm with}\quad{\pmb\xi}\equiv\xi_1,\cdots,\xi_n
\end{equation}
\noindent where $A^\mu$ and $\xi_\mu$ denoting the \textit{affinity} and the \textit{extent of the chemical reaction} $\mu$, respectively, and $T$ is the temperature of the system. $A^\mu$ may be expanded in a Taylor series and, since $A({\pmb\xi}_{eq.})=0$, we finally get
\begin{equation}\label{qf7}
\Delta_IS\simeq\frac{1}{2T}\left(\frac{\partial A^\mu}{\partial{\xi}_\nu}\right)_{eq.}\alpha_\mu\alpha_\nu
\end{equation}
\noindent where the summation convention on the repeated indexes $\mu$ and $\nu$ has been adopted. Using the abbreviation
\begin{equation}\label{qf8}
q^{\mu\nu}\equiv\frac{1}{2T}\left(\frac{\partial A^\mu}{\partial{\xi}_\nu}\right)_{eq.}
\end{equation}
\noindent Eq.~(\ref{qf7}) can be brought into the form
\begin{equation}\label{qf9}
\Delta_IS=q^{\mu\nu}\alpha_\mu\alpha_\nu
\end{equation}
\noindent Notice that expression~(\ref{qf9}) is valid in general and coefficient $q^{\mu\nu}$ depend on the type of fluctuations we are considering \cite{prigogine1}. We also have
\begin{equation}\label{qf10}
\frac{A^\mu}{T}=\frac{1}{T}\left(\frac{\partial A^\mu}{\partial\alpha_\nu}\right)_{eq.}\!\!\!\!\alpha_\nu=q^{\mu\nu}a_\nu=\frac{\partial\Delta_IS}{\partial\alpha_\mu}=X^\mu
\end{equation} 
\noindent Hence,
\begin{equation}\label{qf11}
\alpha_\mu=\widehat{q}_{\mu\nu}X^\mu
\end{equation} 
\noindent with $\widehat{q}_{\mu\nu}$ denoting the \textit{reciprocal matrix} of $q^{\mu\nu}$ i.e., $\widehat{q}_{\nu\kappa}q^{\kappa\mu}=\delta^\mu_\nu$, with $\delta^\mu_\nu$ denoting the Kronecker delta tensor. By taking into account Eq.~(\ref{qf5}) and Eq.~(\ref{of1}), we finally get 
\begin{equation}\label{qf12}
\frac{A^\mu}{T}={\slash\!\!\! k}_{\!{B}}q^{\mu\nu}{\bf K}_\nu\qquad ; \qquad v_\mu={\slash\!\!\! k}_{\!{B}} L_{\mu\rho}q^{\rho\nu}{\bf K}_\nu={\slash\!\!\! k}_{\!{B}}q^{\nu}_\mu{\bf K}_\nu
\end{equation} 
\noindent where $q^\nu_\mu\equiv L_{\mu\rho}q^{\rho\nu}$ and $v_\mu$ is the \textit{rate of the chemical reaction} $\mu$. Expressions~(\ref{qf12}) are of a fairly general nature and, in general, we have
\begin{equation}\label{qf13}
X^\mu={\slash\!\!\! k}_{\!{B}}q^{\mu\nu}{\bf K}_\nu\qquad ; \qquad J_\mu={\slash\!\!\! k}_{\!{B}}q^{\nu}_\mu{\bf K}_\nu
\end{equation} Eqs~(\ref{qf13}) are the expressions for the discretized thermodynamic forces and the thermodynamic fluxes, respectively, and matrix $q_{\mu\nu}$ depends on the nature of the fluctuating parameters.

\subsection{Discretization of the Glansdorff-Prigogine Dissipative Quantity}\label{qgp}
\noindent By taking into account Eq.~(\ref{mep2}), in Onsager's region the rate of the local entropy production strength reads
\begin{equation}\label{qgp1}
\partial_t\sigma=J_\mu\partial_tX^\mu+X^\mu\partial J_\mu= L_{\mu\nu}X^\mu\partial_tX^\nu+L^{\mu\nu}J_\mu\partial_tJ_\nu=2P
\end{equation}
\noindent where $P$ is the Glansdorff-Prigogine dissipative quantity. By performing transformation~(\ref{qep3}) we get
\begin{equation}\label{qgp2}
\partial_t\sigma=\sum_{\mu=1}^n (X^{'\mu}\partial_tX^{'\mu}+J'_\mu\partial_tJ'_\mu)
\end{equation}
\noindent Without loss of generality, we may assume that during the relaxation, $X^{'\mu}$ decreases at a rate proportional to its current value i.e.
\begin{equation}\label{qqgp3}
\partial_tX^{'\mu}=-\lambda X^{'\mu}
\end{equation}
\noindent where $\lambda$ is a positive \textit{rate constant}. Since in Onsager's region $J'_\mu=L_{\mu\nu}X^{'\nu}$, we also have
\begin{equation}\label{qqgp4}
\partial_tJ'_\mu=-\lambda J'^\mu
\end{equation}
\noindent where we have assumed that the transport coefficients are independent of time. So, in terms of the new variables, the rate of the local entropy production strength reads \footnote{Recall that after transformation~(\ref{qep3}) $\sqrt{L}\rightarrow1$.}
\begin{equation}\label{qgp5}
\frac{1}{2}\frac{1}{{\Omega}'}\int_{\Omega'} \partial_t\sigma({\bf X'},t)d^nX'=\frac{1}{{\Omega}'}\int_{\Omega'} P({\bf X'},t)d^nX'=-\frac{\lambda}{2\Omega'}\int_{\Omega'}\sum_{\mu=1}^n\left(\mid X^{'\mu}\mid^2+\mid J'_\mu\mid^2\right)d^nX
\end{equation}
\noindent By following the same procedure we used to discretize the total entropy production strength, we finally get
\begin{equation}\label{qgp6}
{\bf P}\equiv\frac{1}{{\Omega}'}\int_{\Omega'} P({\bf X'},t)d^nX'=-\lambda \sum_{\mu=1}^n\sum_{\bf K} {\slash\!\!\! k}_{\!{B}}\omega_{\bf K}\left({\rm a}_{\bf K}^{(\mu)+}{\rm a}_{\bf K}^{(\mu)}+\frac{1}{2}\right)
\end{equation}
\noindent The eigenvalues ${\mathcal P}$ of the operator ${\bf P}$ are 
\begin{equation}\label{qgp7}
{\mathcal P}=<n_1, n_2,\cdots,n_\kappa,\cdots\mid{\bf P}\mid n_1,n_2,\cdots,n_\kappa,\cdots>=-\lambda \sum_{\mu=1}^n\sum_{\bf K}\left(n^{(\mu)}_\kappa+\frac{1}{2}\right){\slash\!\!\! k}_{\!{B}}\omega_{\bf K}\leq 0
\end{equation}
\noindent where the inequality is saturated at the steady state.

\section{Correspondence with the Einstein-Prigogine Theory of Fluctuations in the Limit of Large Systems}\label{cep}
\noindent The (partial) reconciliation with the classical theory of fluctuations can be obtained by stating the following \textit{correspondence principle}: 

\noindent \textit{The entropy production of the system computed with the formalism of the canonical commutation rules must recover the expression derived by Einstein-Prigogine's fluctuations theory in the limit of macroscopic systems}. 

\noindent We have already noted that, in principle, the constant term 
\begin{equation}\label{cep1}
\sigma_0=\frac{1}{2}\sum_{\mu=1}^n\sum_{\bf K} {\slash\!\!\! k}_{\!{B}}\omega_{\bf K}
\end{equation}
\noindent diverges. In the literature, this kind of divergence is referred to as the \textit{ultraviolet divergence}. More in general, an ultraviolet divergence is a situation in which a sum (or an integral) diverges in physical phenomena at infinitesimal distances. Since an infinite result is unphysical, ultraviolet divergences often require special treatment to remove unphysical effects inherent in the perturbative formalisms \footnote{In this work, we limit ourselves to studying systems in the linear region of thermodynamics assuming that the nonlinear contributions are of lower order.}. In our specific case, this drawback can easily be solved by applying the above-mentioned \textit{correspondence principle}. As previously mentioned, in nonequilibrium systems entropy production arises from the irreversible processes and fluctuations occurring at the microscopic level. Let us then consider a macroscopic system in thermodynamic equilibrium where very small fluctuations around this state occur. According to Einstein-Prigogine's fluctuations theory the probability $P$ of fluctuation $\Delta\xi_\mu$ is proportional to the exponential of the corresponding entropy production (with negative sign)  $-\Delta_IS$ divided by Boltzmann's constant $k_B$ \cite{tolman}, \cite{fowler}, \cite{greene}, \cite{prigogine7}. Hence, the probability of finding a state in which the values of $\alpha_\mu$ lie between $\alpha_\mu$ and $d\alpha_\mu$ is
\begin{equation}\label{cep2}
Pd\alpha_1\cdots d\alpha_n=P_0\exp(-\Delta_IS/k_B)d\alpha_1\cdots d\alpha_n
\end{equation}
\noindent where $P_0$ ensures normalization to unity. Let us now calculate the following average
\begin{equation}\label{cep3}
{\overline{\overline{X^\mu\alpha_\nu}}}=\int\cdots\int X^\mu\alpha_\nu Pd\alpha_1\cdots d\alpha_n
\end{equation}
\noindent It should be immediately specified that \textit{expression~(\ref{cep2}) is valid for small spontaneous fluctuations around the thermodynamic equilibrium only and not for systematic deviations from equilibrium}. Noting that 
\begin{equation}\label{cep4}
X^\mu=\frac{\partial\Delta_IS}{\partial\alpha_\mu}=-k_B\frac{\partial}{\partial\alpha_\mu}\left(\log\frac{P}{P_0}\right)=-\frac{k_B}{P}\frac{\partial P}{\partial\alpha_\mu}
\end{equation}
\noindent we have
\begin{equation}\label{cep5}
{\overline{\overline{X^\mu\alpha_\nu}}}=-k_B\int \frac{\partial P}{\partial\alpha_\mu}\alpha_\nu d\alpha_1\cdots d\alpha_n
\end{equation}
\noindent Partial integration over the $\alpha_i$ gives 
\begin{equation}\label{cep6}
{\overline{\overline{X^\mu\alpha_\nu}}}=k_B\delta^\mu_\nu
\end{equation}
\noindent where we have taken into account that $P$ vanishes at the boundary, i.e. $P=0$ for $\alpha_\nu=\pm\infty$, and that the probability is normalized to unity. Hence, the fluctuation phenomena are accompanied by a variation in entropy production. According to Eq.~(\ref{cep6}) and by taking into account Eq.~(\ref{qf7}), we get
\begin{equation}\label{cep7}
{\overline{\overline{\Delta_IS}}}=\frac{1}{2}{\overline{\overline{X^\mu\alpha_\mu}}}=\frac{n}{2}k_B
\end{equation}
\noindent with $n$ denoting the number of the independent thermodynamic forces. We recall that Eq.~(\ref{cep7}) is valid for \textit{small spontaneous fluctuations around thermodynamic equilibrium only}. Expression~(\ref{cep1}) corresponds to the total entropy production for a macroscopic system generated by very small fluctuations around the thermodynamic equilibrium. Now we calculate the eigenvalues of the ${\pmb\Delta}_{\bf I}{\bf S}$ operator by using the canonical commutation rules. By taking into account Eq.~(\ref{qf9}) and Eq.~(\ref{qf11}), the expression for the ${\pmb\Delta}_{\bf I}{\bf S}$ operator reads
\begin{equation}\label{cep8}
{\pmb\Delta}_{\bf I}{\bf S}=\frac{1}{4\Omega}\int\left({\hat q}_{\mu\nu}X^\mu X^\nu+q^{\mu\nu}\alpha_\mu\alpha_\nu\right)\sqrt{L}\ d^nX
\end{equation}
\noindent Notice that for the second law of thermodynamics ${\hat q}_{\mu\nu}$ is a definite positive matrix (see Eq.~(\ref{qf9})). Therefore, there is a linear coordinates transformation such that
\begin{equation}\label{cep9}
{\alpha'}_\lambda=A^\kappa_\lambda \alpha_\kappa,\quad{\rm with}\ \ A^\mu_\nu\ {\rm such\ that}\ \ A^\alpha_\lambda q^{\lambda\kappa}A^\beta_\kappa={\rm I}^{\alpha\beta}
\end{equation}
\noindent With a suitable definition of a new set of dimensionless variables and following the same procedure adopted in subsection~\ref{qep}., we finally obtain
\begin{equation}\label{cep10}
{\pmb\Delta}_{\bf I}{\bf S}=\sum_{\mu=1}^n\sum_{\bf K}{\slash\!\!\! k}_{\!{B}} \left(\mid {\tilde X}^\mu_{\bf K}\mid^2+\mid {\tilde \alpha}_{\mu,{\bf K}}\mid^2\right)
\end{equation}
\noindent or, introducing the operators of \textit {creation} and \textit {destruction} ${\rm a}_{\bf K}^{(\mu)+}$ and ${\rm a}_{\bf K}^{(\mu)}$, and the number operator ${\bf n}_{\bf K}^{(\mu)}$, we get
\begin{equation}\label{cep11}
{\pmb\Delta}_{\bf I}{\bf S}=\sum_{\mu=1}^n\sum_{\bf K} {\slash\!\!\! k}_{\!{B}}{\bf n}_{\bf K}^{(\mu)}+\frac{n}{2}\sum_{\bf K} {\slash\!\!\! k}_{\!{B}}
\end{equation}
\noindent where the canonical commutation rules have been taken into account. The components of the hyper-vector $K_\mu$ must be compatible with the conditions of periodicity at the boundaries of a hyper-cuboid with sides $l_\mu$. Hence,
\begin{equation}\label{cep11a}
K_\mu=\frac{2\pi m_\mu}{l_\mu}\qquad{\rm with}\quad m_\mu=0, \pm 1,\pm 2,\cdots \qquad \mu=1,\cdots, n
\end{equation}
\noindent where $m_\mu$ is a vector with integer components. Therefore, the eigenvalue $\Delta_IS|_0$ related to the ground state of ${\pmb\Delta}_{\bf I}{\bf S}$ is
\begin{equation}\label{cep12}
\Delta_IS|_0=\frac{n}{2}{\slash\!\!\! k}_{\!{B}}M\qquad {\rm where}\quad M\equiv \Pi_{\mu=1}^n m^{(Max)}_\mu
\end{equation}
\noindent with $m^{(Max)}_\mu$ denoting the maximum number of terms that the vector number $m_\mu$ can assume and $\Pi$ is the product notation, respectively. Eq.~(\ref{cep12}) coincides with the ground state calculated by the Einstein-Prigogine fluctuations theory by setting the following value for $M$ (see Eq.~(\ref{cep7})):
\begin{equation}\label{cep12a}
M=\frac{k_B}{{\slash\!\!\! k}_{\!{B}}}\simeq 8.62\ 10^7
\end{equation} The expression for the operator ${\pmb\Delta}_{\bf I}{\bf S}$ finally reads
\begin{equation}\label{cep13}
{\pmb\Delta}_{\bf I}{\bf S}=\sum_{\mu=1}^n\sum_{\bf K} {\slash\!\!\! k}_{\!{B}}{\bf n}_{\bf K}^{(\mu)}+\frac{n}{2}k_B
\end{equation}
\noindent having discrete eigenvalues
\begin{equation}\label{cep14}
\Delta_IS=\sum_{\mu=1}^n\sum_{\bf K} {\slash\!\!\! k}_{\!{B}}n_{\bf K}^{(\mu)}+\frac{n}{2}k_B
\end{equation}

\subsection{Exemple}\label{ex}
\begin{figure}
\includegraphics[width=6cm]{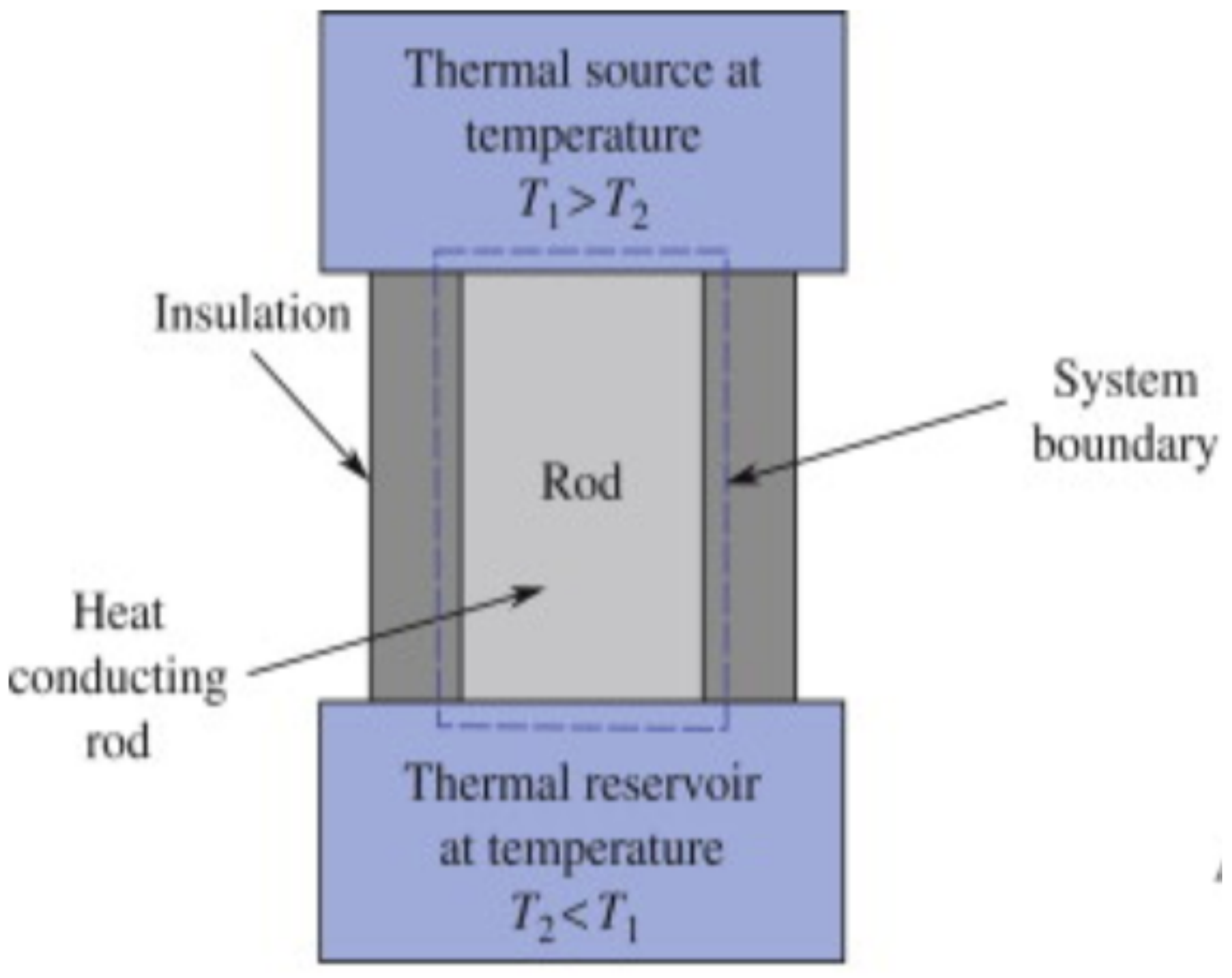}
\caption {\textit{Entropy production due to heat transfer.} A rod of copper is placed between two reservoirs at temperature $T_1$ (the hot one) and $T_2$ (the cold one). The rod is thermally isolated and can exchange heat only at its ends. For  $T_1=10^{\circ}C$, $T_2=0^{\circ}C$, the length of the rod $6cm$, and the cross-sectional area $9cm^2$, after one second the order of magnitude of the number of $TUI$ is about $n^{(1)}\simeq 5.40\ 10^{20}$ and the total number of \textit{TUI} amounts to $Mn^{(1)}\simeq 4.66\ 10^{28}$.}
\label{Example}
\end{figure}

\noindent Consider a copper rod with length $L=6cm$ and cross-section $A=9cm^2$. The rod is placed between two ideal reservoirs $T_1=10^\circ C$ and $T_2=0^\circ C$ and laterally is thermally insulated so that it can only exchange heat at its ends (see Fig.~\ref{Example}). Our goal is to estimate the number of $TUI$ present in the copper rod.
\noindent Denoting with ${\dot Q}$ the heat flow rate  from $T_1$ to $T_2$, the total entropy production strength reads
\begin{equation}\label{ex1}
\sigma_T={\dot Q}\left(\frac{1}{T_2}-\frac{1}{T_1}\right)
\end{equation}
\noindent The heat flow rate flowing in a rod is given by Fourier's law:
\begin{equation}\label{ex2}
{\dot Q}=\kappa_c\frac{A}{L}(T_1-T_2)
\end{equation}
\noindent with $\kappa_c\simeq 385\ W/(m K)$ denoting the thermal conductivity of copper \cite{young}. Hence, the total entropy production strength is
\begin{equation}\label{ex3}
\sigma_T=\kappa_c\frac{A}{L}\frac{(T_1-T_2)^2}{T_1T_2}=7.46\ 10^{-3} W/K
\end{equation}
\noindent To get the order of magnitude of the number of $TUI$, we set $n=1$ in  Eq.~(\ref{cep14}) (as the system is subject to only one thermodynamic force) and, for simplicity, we consider the case where each mode contains the same number $n^{(1)}$ of $TUI$ i.e., 
\begin{equation}\label{ex4}
\sum_{\bf K}{\slash\!\!\! k}_{\!{B}}n_{\bf k}^{(1)}\sim M{\slash\!\!\! k}_{\!{B}}n^{(1)}
\end{equation}
\noindent From  Eq.~(\ref{cep14}) we get
\begin{equation}\label{ex5}
\Delta_IS\simeq M {\slash\!\!\! k}_{\!{B}}n^{(1)}+\frac{1}{2}k_B=k_B\left(n^{(1)}+\frac{1}{2}\right)
\end{equation}
\noindent Comparing Eq.~(\ref{ex5}) with Eq.~(\ref{ex3}), we obtain
\begin{equation}\label{ex6}
n^{(1)}=\frac{\sigma_T}{k_B}\Delta t-\frac{1}{2}
\end{equation}
\noindent If we apply typical "{\textit{human-scale}" values, $\Delta t\sim1\ sec.$, then $n^{(1)}\simeq 5.50\ 10^{20}$, and the total number of \textit{TUI} amounts to $Mn^{(1)}\simeq 4.66\ 10^{28}$. These are very big numbers, so the system is indeed within the correspondence limit. We can easily also check why we perceive a continuum of energy production strength in this limit. With $\omega=1\ rad/sec$, the difference between two discretized entropy production strengths is ${\slash\!\!\! k}_{\!{B}}\omega\simeq1.6\ 10^{-31}$ W/K, well below what we normally resolve for macroscopic systems. One then describes this system through an emergent classical limit. Notice that Eq.~(\ref{ex6}) tells us that we have no $TUI$ for
\begin{equation}\label{ex7}
\sigma_T\Delta t=\frac{1}{2}k_B
\end{equation}
\noindent This result is in line with our expectations as, for the Einstein-Prigogine fluctuations theory, the ground state of the entropy production is $1/2k_B$ and, in accordance with our formalism, this corresponds to the vacuum state of the entropy production strength $|0,\cdots,0>$, i.e., no $TUI$.

\section{Microscopic Reversibility and the Onsager Reciprocity Relations}\label{orr}
\noindent The transport flux-force relation given by Eq.~(\ref{of1}) involves the transport coefficients $\tau_{0\mu\nu}$ where the symmetric coefficients of this matrix, denoted by $L_{\mu\nu}$, are called \textit{phenomenological coefficients}. An important theorem due to Onsager states that \cite{onsager1}, \cite{onsager2}  
\begin{equation}\label{orr1}
L_{\mu\nu}=L_{\nu\mu}
\end{equation}
\noindent These relations, referred to as \textit{Onsager's reciprocity relations}, express that when the flux, corresponding to the irreversible processes $\mu$, is influenced by the thermodynamic forces $X^\nu$ of the irreversible processes $J_\nu$ through the interference $L_{\nu\mu}$, then the flux $J_\nu$ is also influenced by the thermodynamic forces $X^\mu$, through the same interference $L_{\mu\nu}$ i.e., $L_{\mu\nu}=L_{\nu\mu}$. Prigogine showed the validity of the Onsager reciprocity relations, starting from the property of microscopic reversibility and by using the Einstein-Prigogine fluctuations theory \cite{prigogine1}. However, his proof is valid only in case of (very) small fluctuations around the ground state i.e., around the thermodynamic equilibrium or the local non-equilibrium state. In this section, we shall demonstrate that, under the assumption of the microscopic reversibility property, the formalism of the canonical commutation rules preserves the validity of the Onsager reciprocity relations in the whole linear region of thermodynamics (and not only in the region very close to the ground state). For \textit{Microscopic reversibility} we mean the symmetry of all mechanical equations of motion of single particles with respect to time. Let us consider a fluctuation $\alpha_\mu({\bf X},t)$ in ${\bf X}$ at time $t$ and a fluctuation $\alpha_\nu({\bf X},t+\tau)$ in ${\bf X}$ after a time interval $\tau$, and form the average (in the thermodynamic space) of the product of both quantities:
\begin{equation}\label{orr2}
\frac{1}{2\Omega}\int\left(\alpha_\mu ({\bf X},t)\alpha_\nu ({\bf X},t+\tau)+\alpha_\nu ({\bf X},t+\tau)\alpha_\mu ({\bf X},t)\right)\sqrt{L}d^nX
\end{equation}
\noindent We also consider the value of the average product 
\begin{equation}\label{orr3}
\frac{1}{2\Omega}\int\left(\alpha_\nu ({\bf X},t)\alpha_\mu ({\bf X},t+\tau)+\alpha_\mu ({\bf X},t+\tau)\alpha_\nu ({\bf X},t)\right)\sqrt{L}d^nX
\end{equation}
\noindent where fluctuation $\alpha_\mu ({\bf X},t+\tau)$ occurs after the time interval $\tau$. The two expressions (\ref{orr2}) and (\ref{orr3}) differ only in the temporal order of the two fluctuations or, say in other words, by substitution $t\rightarrow -t$. So, under the assumption of microscopic reversibility, we get
\begin{align}\label{orr4}
&\frac{1}{2\Omega}\int\left(\alpha_\mu ({\bf X},t)\alpha_\nu ({\bf X},t+\tau)+\alpha_\nu ({\bf X},t+\tau)\alpha_\mu ({\bf X},t)\right)\sqrt{L}d^nX=\\
&\frac{1}{2\Omega}\int\left(\alpha_\nu ({\bf X},t)\alpha_\mu ({\bf X},t+\tau)+\alpha_\mu ({\bf X},t+\tau)\alpha_\nu ({\bf X},t)\right)\sqrt{L} d^nX\nonumber
\end{align}
\noindent Subtracting the same quantity 
\begin{equation}\label{orr5}
\frac{1}{2\Omega}\int\left(\alpha_\nu ({\bf X},t)\alpha_\mu ({\bf X},t)+\alpha_\mu ({\bf X},t)\alpha_\nu ({\bf X},t)\right)\sqrt{L}d^nX
\end{equation}
\noindent from both members of Eq.~(\ref{orr4}) and dividing the expressions by $\tau$, as $\tau\rightarrow 0$ we get 
\begin{equation}\label{orr6}
\frac{1}{2\Omega}\int\left(\alpha_\mu{\dot\alpha}_\nu+{\dot\alpha}_\nu\alpha_\mu\right)\sqrt{L}d^nX=\frac{1}{2\Omega}\int\left(\alpha_\nu{\dot\alpha}_\mu+{\dot\alpha}_\mu\alpha_\nu\right)\sqrt{L} d^nX
\end{equation}
\noindent or
\begin{equation}\label{orr7}
\frac{1}{2\Omega}\int\left(\alpha_\mu J_\nu+J_\nu\alpha_\mu\right)\sqrt{L}d^nX=\frac{1}{2\Omega}\int\left(\alpha_\nu J_\mu+J_\mu \alpha_\nu\right)\sqrt{L}d^nX
\end{equation}
\noindent By applying the flux-force linear transport relations $J_\mu=L_{\mu\nu}X^\nu$ we find
\begin{equation}\label{orr8}
\L_{\nu\kappa}\frac{1}{2\Omega}\int\left(\alpha_\mu X^\kappa+X^\kappa\alpha_\mu \right)d^nX=L_{\mu\kappa}\frac{1}{2\Omega}\int\left(\alpha_\nu X^\kappa+X^\kappa\alpha_\nu \right)d^nX
\end{equation}
\noindent As for the integrals, if $\kappa=\mu$ or $\kappa=\nu$ we get operator ${\pmb\Delta}_{\bf I}{\bf S}$ given by Eq.~(\ref{cep13}). However, their eigenvalues vanish for $\kappa\neq\mu$ or $\kappa\neq\nu$ (all commutators $[a_\mu,a_{\nu}]=[a_\mu,a^+_{\nu}]=[a^+_\mu,a^+_{\nu}]=0$ for $\mu\neq\nu$). Hence, in terms of eigenvalues, Eq.~(\ref{orr8}) gives
\begin{equation}\label{orr9}
2L_{\nu\kappa}\delta_\mu^\kappa\Delta_IS=2L_{\mu\kappa}\delta_\nu^\kappa\Delta_IS
\end{equation}
\noindent or
\begin{equation}\label{orr10}
L_{\nu\mu}=L_{\mu\nu}
\end{equation} i.e., the Onsager relation which we sought. To sum up, Prigogine in \cite{prigogine1} showed the validity of Onsager's reciprocity relation for small spontaneous fluctuations around the thermodynamic equilibrium only. Here, \textit{we demonstrated that Onsager's relations are satisfied even for systematic deviations from the ground state, as long as the flux-force transport relations remain linear}. 

\noindent Notice that having checked the validity of Onsager's reciprocity relations over the whole range of linear thermodynamics through the canonical commutation relations (ccr) is not a mere mathematical exercise. It is a very important result as it demonstrates that the ccr can only be formulated as given by Eq.~(\ref{ccr1}).

\section{Effect of the Volume of the System on the Simultaneous Measurement of Entropy Production Strength and Time}\label{vccr}
 
\noindent In terms of the local thermodynamic variables ${\bar\sigma}$, ${\bar X}_{\bf K}^\mu$, and ${\bar\alpha}_{\mu ,{\bf K}}$, the uncertainty relations read
\begin{align}\label{vccr1}
&[t,{\bar\sigma}]=i\frac{{\slash\!\!\! k}_{\!{B}}}{2V}\\
&[{\bar\alpha}_{\mu ,{\bf K}},{\bar X}_{\bf K'}^\nu]=i\frac{{\slash\!\!\! k}_{\!{B}}}{2V}\delta_{\mu\nu}\delta_{{\bf K}{\bf K'}}\nonumber
\end{align}
\noindent i.e., the uncertainty relations increase as the volume of the system decreases. This result is in line with our expectations: as the volume decreases and we move to the mesoscopic level, the simultaneous measurement of entropy production strength and time becomes more indeterminate. As already remarked, this is due to our increased ability to see more and more details of collisions and to the influence of fluctuations. These are the effects that cause the inherent uncertainty associated with the smallest systems. The transport coefficients (and then the entropy production) arise from particle collisions, so discrete behaviors may be observed on a smaller scale. Fluctuations arise due to the probabilistic nature of individual particle interactions and movements. Fluctuations become more significant as we move to the mesoscopic scale, where systems are smaller and composed of a limited number of particles. In smaller systems, these fluctuations have a greater impact on the system's overall behavior as they can lead to transient changes in entropy production strength. Summing up, the limited number of particles in the system leads to a greater sensitivity to individual events, making it more challenging to precisely determine the entropy production strength at any specific moment. 

\section{Conclusions and Future Works}\label{cfw}
\noindent The physical world can exhibit different behaviors and be described by different laws depending on the scale of observation. The laws governing macroscopic systems may not necessarily apply directly to mesoscopic or microscopic systems, and vice versa. Overall, the physical world is governed by a hierarchy of laws that apply at different scales, ranging from classical mechanics and thermodynamics at the macroscopic scale to quantum mechanics and emergent behavior at the mesoscopic and microscopic scales. Understanding these scale-dependent laws is crucial for comprehending the diverse range of phenomena observed in the universe. Emergent phenomena, as seen at the mesoscopic scale, can arise due to the collective behavior of a large number of particles. These phenomena may not have direct counterparts in either classical or quantum mechanics alone, and they often require new theoretical frameworks or models to describe them. One of the main objectives of the \textit{Brussels School of Thermodynamics}, directed by Prof. Ilya Prigogine, was to investigate systems on a mesoscopic scale with the aim to discover the fundamental laws governing them. In this work, inspired by this goal and encouraged by recent experimental results, we established the canonical commutation rules (ccr) valid at the mesoscopic scale. The ccr~(\ref{vccr1}) shows that the closer we get to the mesoscopic level, the more indeterminate becomes the simultaneous measurement of the canonically conjugate variables. For example, the uncertainties $\Delta\sigma$ and $\Delta t$ in simultaneously existing values of the entropy production strength and time are related by the expression $\Delta t\Delta\sigma\geq {\slash\!\!\! k}_{\!{B}}/2$; the greater the accuracy with which one of these quantities is measured, the less the accuracy with which the other can be measured at the same time. We have seen that fundamental quantities such as the total entropy production, the thermodynamic variables conjugate to the thermodynamic forces, and the Glansdorff-Prigogine's dissipative variable are discretized at the mesoscopic scale. The \textit{ultraviolet divergence problem} has been solved by applying the \textit{correspondence principle} to Einstein-Prigogine's fluctuations theory in the limit of macroscopic systems. Incidentally, we have also shown that the formalism based on the canonical commutation rules confirms the validity of the Onsager reciprocity relations over the entire linear region of thermodynamics. 

\noindent \noindent Future works are devoted to the discretization of the total entropy production, the thermodynamic variables conjugate to the thermodynamic forces, and the Glansdorff-Prigogine’s dissipative variable for systems out of Onsager's region.  This task will be accomplished by using the field theory developed in \cite{sonnino3}, \cite{sonnino7}, and \cite{sonnino8}. This will involve, in the expression for the entropy production, terms not only quadratic but cubic or even of a higher degree i.e., terms, for example, of the type ($a + a^+)^4$, etc. The new expression for the total entropy production will then be the sum of two terms: the expression found for systems in Onsager's region and a sort of \textit{interaction term} (which does not commute with the previous one) multiplied by a coupling constant $\epsilon$. The discretized entropy production is obtained by a perturbative expansion by assuming that the constant $\epsilon$ is small enough.

\section{Acknowledgments}
Preliminary discussions on this matter were held with the late Prof. Enrique Tirapegui. I would like to pay a special tribute to his memory. Enrique always supported and encouraged me to pursue my research in the field of Thermodynamics of Irreversible Processes. I would also like to pay tribute to Prof. Ilya Prigogine. My strong interest in this domain of research is due to him, who promoted the \textit {Brussels School of Thermodynamics} at the Universit{\'e} Libre de Bruxelles (U.L.B.), where I took my doctorate in Physics. I am grateful to Prof. Pasquale Nardone and Dr Philippe Peeters, from the U.L.B., and Prof. Carlo Maria Becchi of the University of Genoa (Italy) for the useful discussions and suggestions. 



\end{document}